\RequirePackage{fix-cm}
\documentclass[smallextended]{svjour3}       
\smartqed  
\usepackage{times}
\usepackage{subfigure}
\usepackage{fancyheadings}
\usepackage[cp1251]{inputenc}
\usepackage[english]{babel}
\usepackage{graphicx}
\usepackage{color}
\usepackage{cite}
\usepackage{amsmath,amssymb,enumerate,bbm,color,verbatim,setspace}
\usepackage{mathptmx}

\newtheorem{rem}{Remark}[section]

\begin{document}

\title{Unnormalized quasi-distributions and tomograms \\of quantum states
}


\author{V.I. Man'ko         \and
        L.A. Markovich 
}


\institute{V.I. Man'ko \at
              P. N. Lebedev Physical Institute, Russian Academy of Sciences\\
Leninskii Prospect 53, Moscow 119991, Russia \\
Moscow Institute of Physics and Technology\\
 Institutskii Per. 9, Dolgoprudny Moscow Region 141700, Russia
  \email{manko@lebedev.ru}                  
           \and
            L.A. Markovich\at
 Institute for information transmission problems, Moscow,\\ Bolshoy Karetny per. 19, build. 1, Moscow 127051, Russia\\
V. A. Trapeznikov Institute of Control Sciences, Moscow,\\Profsoyuznaya 65, 117997 Moscow, Russia\\
Moscow Institute of Physics and Technology\\
 Institutskii Per. 9, Dolgoprudny Moscow Region 141700, Russia\\
 \email{kimo1@mail.ru}
}

\date{Received: date / Accepted: date}

\maketitle

\begin{abstract}
Tomograms and quasi-distribution functions like Wigner,  Glauber - Sudarshan $P$- and Husimi $Q$- functions that violate the standard normalization condition are considered.
Conditions under which a reconstruction of the density matrix using these tomograms and quasi-distribution functions is possible are obtained. Three different examples of states like the de Broglie plane wave, the Moschinsky shutter problem and the stationary state  of the charged particle in the uniform and constant electric field are studied. Their tomograms and quasi-distribution functions expressed in terms of the Dirac delta function, the Airy function and the  Fresnel integrals  are shown to violate the standard normalization condition and thus the density matrix of the state can not always be reconstructed.
\end{abstract}

\section{Introduction}
\label{intro}
\par In literature there are different formulations of quantum mechanics like the wavefunction, matrix, path integral, phase space, density matrix, second quantization, variational, pilot wave, Hamilton–Jacobi formulations and etc. However, one of the usual ways to describe a quantum mechanical system in the phase space is to use the so-called Wigner
distribution $W(q,p)$ (cf. \cite{Wigner}). For a given state $|\psi\rangle$ the density matrix operator $\hat{\rho}=|\psi\rangle\langle\psi|$ can be constructed and expressed
in the position $q$ and the momentum $p$ representations. Thus, the Wigner distribution can be interpreted as an intermediate representation between
this two. It is known that the true probability $F_{Q}(q,p)$ has to fulfill  the probabilities axioms
\begin{eqnarray}F_{Q}(q,p)\geq0,\label{21_0}\end{eqnarray}
\begin{eqnarray}\iint\limits_{-\infty}^{\infty}F_{Q}(q,p)dqdp=1.\label{21_01}\end{eqnarray}
However, the Wigner function is a quasi-distribution function because it does not satisfy, in general, the
condition of the probability function, namely  it can  take negative values. There are some other quasi-distribution functions which correspond to the density matrix in different representations like  Glauber-Sudarshan $P$-function (cf. \cite{Glauber,Sudarshan}) and Husimi $Q$-function (cf. \cite{Husimi}). These functions can be useful in many applications. For example, the Husimi $Q$-function is used  for nucleon tomography in \cite{Hagiwaraa} and it has numerous applications in statistical physics, condensed matter physics, quantum optics, quantum chaos, and also in atomic and nuclear physics (cf. \cite{Lee,Kunihiro}).
\par Recently the new formulation of quantum mechanics, namely the probability representation of the quantum mechanics was introduced in \cite{ManciniMankoTombesi,ManciniMankoTombesi2}.
To describe quantum states a special function, that is a fair probability distribution, was suggested. This function called a symplectic tomogram is related to the Wigner function
of quantum states by means of integral Radon transform (e.g. \cite{Radon}). Unlike the Wigner function,  the tomogram satisfies all criteria of the distribution function \eqref{21_0} and \eqref{21_01}. The tomogram and all  quasi-distribution functions and  are related to each other and to the density matrix operator by one-to-one transformations, but only
the tomogram has all specific properties of the standard probability distribution and thus completely describes the quantum state.
\par However there are such quantum states for which neither the  tomogram nor the Wigner function
and the $P$- and $Q$-functions violate the standard normalization condition \eqref{21_01}. As an example the de Broglie plane wave can be named. The violation of the standard normalization condition can be due to  several reasons. First, the function can depend only on the part of its variables,
i.e. $W(q,p)\equiv W(q)$. Second, the function can be of a specific form or it may contain some special functions, like, for example, the Dirac delta function, the Airy function or the Fresnel integral.
 \par The aim of the paper is to study how the violation of the standard normality condition \eqref{21_01} impacts the tomogram and the quasi-probabilities.
 To this end several special states are considered. First we study the de Broglie plane wave in the momentum and coordinate forms whose Wigner functions are known to be the delta functions depended only on one of its variables $p$ or $q$, respectively. In this case the standard Radon transform to obtain the tomogram is not applicable.
 Authors develop explicit tomogram formulas  and their suitability for the reconstruction of the density matrix are studied for both cases. Next, the Moschinsky shutter problem is considered  (e.g. \cite{Moshinsky}). The Wigner function and the tomogram  depend in this case on the Fresnel integral (cf. \cite{MankoMoshinskySharma}) and do not
 satisfy  condition  \eqref{21_01}. The stationary state  of the charged particle in the uniform and constant electric field is studied too. Its wave
 function is given by the Airy function of the first order and both the Wigner function and the tomogram violate the normalization condition  \eqref{21_01}.
 \par The paper is organized as follows. In  Section~\ref{sec:21_1} we briefly recall the Weyl and the tomographic symbols theory. The transformations between the Wigner,
 the tomogram and the $Q$-function are shown. In Section~\ref{sec:21_2} tree special states whose Wigner function and the tomogram violate the standard
 normalization condition are studied in details.
\section{Weyl and tomographic symbols}\label{sec:21_1}
\par Here, we give a brief review of the quantization procedure. Among the known quantization methods we select the star-product of the operator symbols (e.g. \cite{Bajen,MankoMankoMarmo2000}). We consider an operator $\hat{A}$  acting on a Hilbert space $\mathcal{H}$. The symbol  $f_{\hat{A}}(\overline{x})$ of the operator  $\hat{A}$ is determined by \begin{eqnarray*}f_{\hat{A}}(\overline{x})=Tr\left(\hat{A}\hat{U}(\overline{x})\right),\quad \hat{A}=\int f_{\hat{A}}(\overline{x})\hat{D}(\overline{x})d\overline{x},\end{eqnarray*}
where the set of operators $\hat{U}(\overline{x})$ and their dual set of operators $\hat{D}(\overline{x})$, $\overline{x}=(x_1,x_2,\ldots,x_N)$ were defined (e.g. in \cite{MankoMarmoStornaiolo}). Using the latter operators the star-product can be introduced by
\begin{eqnarray*}f_{\hat{A}\hat{B}}(\overline{x})=f_{\hat{A}}(\overline{x})\ast f_{\hat{B}}(\overline{x})=Tr\left(\hat{A}\hat{B}\hat{U}(\overline{x})\right).\end{eqnarray*}
In our study we use two types of symbols associated with operators, ss. a Weyl symbol and a tomographic symbol. It is known that for the state density operator the Weyl
symbol is nothing else but the Wigner function.
To define the Weyl symbol $W_{\hat{A}}(q,p)$ the following operators are introduced
\begin{eqnarray*}\hat{U}(\overline{x})=\hat{U}(x_1,x_2),\quad \hat{D}(\overline{x})=\hat{D}(x_1,x_2),\quad x_1=q/\sqrt{2}, \quad x_2=p/\sqrt{2}.\end{eqnarray*}
\par Hence we consider  two-dimensional phase space $\mathbb{R}^2$ with coordinate $q$ and momentum $p$.
The bosonic creation and annihilation operators $a$ and $a^{\dagger}$ are  defined on the Hilbert space $\mathcal{H}$ as following
\begin{eqnarray*}\hat{a}=(\hat{q}+i\hat{p})/\sqrt{2},\quad \hat{a}^{\dag}=(\hat{q}-i\hat{p})/\sqrt{2},\end{eqnarray*}
which satisfy the canonical commutation relation $[\hat{a},\hat{a}^{\dag}]=1$. Two families of operators are introduced
\begin{eqnarray*}\hat{U}(q,p)=2\hat{D}(\alpha)(-1)^{a^{\dagger}a}\hat{D}(-\alpha),\quad \alpha=(q+ip)/2,\end{eqnarray*}
\begin{eqnarray*}\hat{D}(q,p)=\frac{1}{\pi}\hat{D}(\alpha)(-1)^{a^{\dagger}a}\hat{D}(-\alpha),\end{eqnarray*}
where the operator $\hat{D}(\alpha)$ is  a unitary displacement operator
\begin{eqnarray*}\hat{D}(\alpha)=\exp\{\alpha a^{\dagger}-\alpha^{\ast}a\}.\end{eqnarray*}
The operators $\hat{q}$ and $\hat{p}$ are the position and the momentum operators, respectively, given by the standard relations
\begin{eqnarray*}\hat{q}\psi(x)=x\psi(x),\quad \hat{p}\psi(x)=-i\hbar\frac{\partial \psi(x)}{\partial x}.\end{eqnarray*}
Finally, the Weyl symbol  $W_{\hat{A}}(q,p)$  of the operator $\hat{A}$ is determined by
\begin{eqnarray}W_{\hat{A}}=2Tr\left(\hat{A}\hat{D}(\alpha)(-1)^{a^{\dagger}a}\hat{D}(-\alpha)\right).\label{21_1}\end{eqnarray}
If we consider a quantum system associated with the density operator \begin{eqnarray*}\widehat{\rho}=|\psi\rangle\langle\psi|,\end{eqnarray*}
that is a positive acting on a Hilbert space $\mathcal{H}$ operator with unit trace, the relation \eqref{21_1}
defines the Wigner function of the state. For a pure generic quantum state $|\psi(t)\rangle$, the Wigner distribution $(\hbar=1)$ is defined as
\begin{eqnarray*}W(q,p,t)&=\int\limits_{-\infty}^{\infty} e^{-ipx}\langle\psi(t)|q-x/2\rangle\langle q+x/2|\psi(t)\rangle dx\\
&=\int\limits_{-\infty}^{\infty} e^{-ipx}\langle q+x/2|\hat{\rho}(t)|q-x/2\rangle dx,\end{eqnarray*}
where $\hat{\rho}(t)=|\psi(t)\rangle\langle\psi(t)|$  is the density matrix of the density operator $\widehat{\rho}$. The Wigner  function satisfies the standard normalization condition
\begin{eqnarray}\int_{-\infty}^{\infty}W(q,p,t)\frac{dpdq}{2\pi}=1\label{21_19}\end{eqnarray}
and
\begin{eqnarray*}\int_{-\infty}^{\infty}W(q,p,t)\frac{dq}{2\pi}=|\langle\psi(t)|p\rangle|^2,\quad\int_{-\infty}^{\infty}W(q,p,t)\frac{dp}{2\pi}=|\langle\psi(t)|q\rangle|^2,\end{eqnarray*}
hold. Another feature of the Wigner function is that it is always real valued. However, it  is not positive-defined and can not be interpreted as the probability distribution at the phase space $(q,p)$. Thus, the Wigner function is called a quasi-distribution function.
\par Now we turn our attention to the tomographic symbol. Two following families of operators are introduced
\begin{eqnarray*}\hat{U}(X,\mu,\nu)=\delta(X\hat{1}-\mu \hat{q}-\nu \hat{p}),\quad \hat{D}(X,\mu,\nu)=\frac{1}{2\pi}e^{iX\hat{1}}e^{-i(\mu \hat{q}+\nu \hat{p})},\end{eqnarray*}
where $X$, $\mu$ and $\nu$ are real variables.
In \cite{MankoMankoMarmoVitale2007} the tomographic symbol of the operator $A$ called the tomogram is defined by the
relation
\begin{eqnarray*}\mathcal{W}_{\hat{A}}(X,\mu,\nu)=Tr\left(\hat{A}\delta(X-\mu \hat{q}-\nu \hat{p})\right).\end{eqnarray*}
Obviously, the  inverse relation reads as
\begin{eqnarray}\hat{A}=\frac{1}{2\pi}\iiint\limits_{-\infty}^{\infty}\mathcal{W}_{\hat{A}}(X,\mu,\nu)e^{i(X-\mu \hat{q}-\nu \hat{p})}dXd\mu d\nu.\label{21_7}\end{eqnarray}
Thus, the symplectic tomogram of the quantum state with the density operator $\hat{\rho}$ is defined as
\begin{eqnarray*}\mathcal{W}(X,\mu,\nu)=Tr\left(\hat{\rho}\delta(X-\mu \hat{q}-\nu \hat{p})\right),\end{eqnarray*}
where the random coordinate $X$ corresponds to the particle's position and the real parameters $\mu$ and $\nu$ label
the reference frame in the classical phase space in which the position is measured.
In contrast to the Wigner function, the tomogram has all the properties of the distribution function. In particular the normalization condition
\begin{eqnarray}\int\limits_{-\infty}^{\infty}\mathcal{W}(X,\mu,\nu)dX=1\label{21_20}\end{eqnarray}
holds for all $\mu$ and $\nu$.   Thus, the symplectic tomogram is the probability distribution of random value $X$ measured in the reference frame in the face space. The latter reference frame is defined by $\mu=s\cos{\theta}$ and $\nu=s^{-1}\sin{\theta}$, where $\theta$ is the rotation angle of the frame axes and $s$ is a scaling parameter.
\par Finally, we recall definitions of  $P$- and $Q$- functions, two most frequently used quasi-distributions.
It is known that the coherent states $|\alpha\rangle\in \mathcal{H}$, that are the normalized eigenkets of the lowering operator $\hat{a}|\alpha\rangle=\alpha |\alpha\rangle$, provide the continuous basis for Hilbert space
\begin{eqnarray*}|\alpha\rangle=e^{-|\alpha|^2/2}\sum\limits_{n=0}^{\infty}\frac{\alpha^n}{\sqrt{n!}}|n\rangle\end{eqnarray*}
In \cite{Husimi} the Husimi $Q$ - function is introduced as
  \begin{eqnarray}Q(\alpha,\alpha')=\langle \alpha|\hat{\rho}|\alpha\rangle\label{21_6}\end{eqnarray}
It is a trace of the density matrix over the basis of coherent states $\{|\alpha\rangle\}$. The $Q$-function can be represented via the Wigner function as
 \begin{eqnarray}
Q(\alpha,\alpha')=\frac{2}{\pi}\int d^2\beta e^{-2|\alpha-\beta|^2}W(\beta,\beta'), \label{21_8}\end{eqnarray}
where
\begin{eqnarray*}&&\int d^2\beta=\int\limits_{-\infty}^{\infty}d\operatorname{Re}\beta\int\limits_{-\infty}^{\infty}d\operatorname{Im}\beta.\end{eqnarray*}
Since \eqref{21_7}, the Husimi distribution is  positive-semidefinite and the normalization condition
\begin{eqnarray*}\iint\limits_{-\infty}^{\infty}Q(q,p,t)\frac{dpdq}{2\pi}=1,\end{eqnarray*}
holds. The Glauber-Sudarshan $P$ - function (cf. \cite{Cahill,Glauber2}) is the expectation value of the normal-ordered $\delta$ operator
 \begin{eqnarray}P(\alpha,\alpha')=\operatorname{Tr}[\hat{\rho}\delta(\alpha^{\ast}- \hat{a}^{\dag})\delta(\alpha- \hat{a})],\label{21_38}\end{eqnarray}
where
 \begin{eqnarray*}&&\delta(\alpha^{\ast}- \hat{a}^{\dag})\delta(\alpha- \hat{a})=\int\frac{d^2c}{\pi}e^{ic(\alpha^{\ast}- \hat{a}^{\dag})}e^{ic^{\ast}(\alpha- \hat{a})}.\end{eqnarray*}
 The Wigner function is the Gaussian convolution of the $P$ - function of the density matrix, i.e.
 \begin{eqnarray*}&&W(\alpha,\alpha')=\frac{2}{\pi}\int d^2\beta e^{-2|\alpha-\beta|^2} P(\beta,\beta').\end{eqnarray*}
Thus, all quasi-distribution functions are  interrelated by the convolution by Gaussian functions. In the case of violation of the standard normalization condition \eqref{21_01} by the Wigner function, the normalization conditions of other functions are also violated.
\subsection{Wigner function and tomogram  as a tool to reconstruct the density matrix}
Here we briefly consider relations between the tomographic and the Weyl symbols. To this end the Radon transform of  the Wigner function
is used (see \cite{Wigner}). As we mentioned in the previous section, the Wigner function is the Weyl
symbol of the von Neumann density matrix $\rho$. By definition, the Wigner function is expressed in terms of density matrix as
\begin{eqnarray}W(q,p)=\int\limits_{-\infty}^{\infty}\rho\left( q+\frac{u}{2},q-\frac{u}{2}\right) e^{-ipu}du.\label{21_2}\end{eqnarray}
The inverse transform is the following
\begin{eqnarray}\rho(x,x')=\frac{1}{2\pi}\int\limits_{-\infty}^{\infty} W\left( \frac{x+x'}{2},p\right) e^{ip(x-x')}dp.\label{21_3}\end{eqnarray}
Let $f(x,y)$ be a continuous function of real variables $x_i\in R^{1}$, $i=1,2$ and it is decreasing sufficiently fast at infinity. The Radon transform is defined as (see \cite{Gelfand})   \begin{eqnarray}R(r,\theta)f(x,y)=\iint\limits_{-\infty}^{\infty}f(x,y)\delta(r-x\cos{\theta}-y\sin{\theta})dxdy,\label{21_16}\end{eqnarray}
   where $r$ is the perpendicular distance from a line to the origin and $\theta$ is the angle formed by the distance vector.
The Radon transform of the Wigner function reads ($\hbar=1$) (see \cite{ManciniMankoTombesi})
   \begin{eqnarray}\mathcal{W}(X,\mu,\nu)=\iint\limits_{-\infty}^{\infty} W(q,p)\delta(X-\mu q-\nu p)\frac{dqdp}{2\pi}.\label{21_17}\end{eqnarray}
To simplify the latter integral we use the notion of Dirac delta function. Often the delta function is simply defined as
\begin{equation}
    \begin{matrix}
    \delta(x) & =
    & \left\{
    \begin{matrix}
    0 & x\neq0\\
   \infty &  x=0
    \end{matrix} \right.
    \end{matrix}.\label{21_40}
\end{equation}
   Since the integral representation of the delta function is
 \begin{eqnarray}\delta(t)=\frac{1}{2\pi}\int\limits_{-\infty}^{\infty}e^{i\omega t}d\omega,\label{21_41}\end{eqnarray}
one can write
   \begin{eqnarray}\mathcal{W}(X,\mu,\nu)=\iint\limits_{-\infty}^{\infty}\left(\int e^{ik(X-\mu q-\nu p)}dk\right)W(q,p)\frac{dqdp}{(2\pi)^2},\label{21_4}\end{eqnarray}
 where $X$, $\mu$, $\nu$ are real numbers. However, for the probability distribution theory the definition \eqref{21_40} is not strict enough and may lead to significant errors. The delta function must be defined as a function satisfying the following relations
\begin{eqnarray}&&\int\limits_{-\infty}^{\infty}f(x) \delta(x)dx=\int\limits_{-\epsilon}^{\epsilon}f(x) \delta(x)dx=f(0),\quad \int\limits_{-\infty}^{\infty} \delta(x)dx=1.\label{21_28}\end{eqnarray}
The latter function  does not exist in the usual sense of a function,
since it is zero everywhere except at a point and thus it is not well defined. However one can find such sequence of  functions those approach a sifting property in a certain limit, e.g.,the sequence of the top hat functions
\begin{equation*}
    \begin{matrix}
    \delta^{(1)}_{n}(x) & =
    & \left\{
    \begin{matrix}
    0 & x<-\frac{1}{n}\\
    \frac{n}{2} & -\frac{1}{n}<x<\frac{1}{n}\\
   0 &  \frac{1}{n}<x
    \end{matrix} \right.
    \end{matrix}.
\end{equation*}
and the following frequently used sequence
\begin{eqnarray*} && \delta^{(2)}_{n}(x)=\frac{n}{\sqrt{\pi}}e^{-n^2x^2}.\end{eqnarray*}
The latter functions satisfy the sifting property \eqref{21_28} in the limit. Thus the delta function only has meaning beneath the integral
sign.
\par The function $f(x)$ in \eqref{21_28} is called a test function and must be defined on a special class of functions that satisfy the sifting property.
This means that the sequence of integrals must converge for $f(x)$ within a class of test functions.
To this end, the delta function is defined as a distribution, also known as a generalized function, which is an object that acts on the class of test functions. In contrast to a usual  function, the distribution function  is not defined in terms of values at points.
\par
It should be stressed that in the conventional mathematical theory of distributions, the sifting property is a priori only defined if $f(x)$ is a test-function (e.g. \cite{Gelfand}). In particular, it is not mathematically rigorous to use \eqref{21_28} where instead of $f(x)$ the delta function (the distribution) is substituted.
According to the no-go theorem by L. Schwartz (e.g. \cite{Schwartz}) it is impossible to define a product of two distributions in such a way that they form an algebra with acceptable topological properties. What is possible is to define the product of distributions when their wave front sets do not meet (e.g. \cite{Hormander}).
 For two delta functions the product can be defined in $\mathbb{R}^2$ as $\delta_{x=0}=\delta_{x_1=0}\cdot\delta_{x_2=0}$ (e.g. \cite{McKay}).
However, multiplying two distributions with the same variable together, namely $\delta^2(x)$,  has no meaning and should be avoided.
\par Comparing expressions \eqref{21_17} and \eqref{21_28}, one can see that the role of the test function $f(x)$ in Radon transform is played by the Wigner function $W(q,p)$.
Since typically the space of test functions consists of all smooth functions on $\mathbb{R}$ with compact support that have as many derivatives as required,
 the Wigner function must satisfy these properties.  However, as it will be shown in the next section in the case of a plane wave, the Wigner function is equal to the delta function.
For such Wigner function, the Radon transformation in its classical form is not applicable and therefore the tomogram can not be determined in this way
 \par Using the inverse Radon relation one can find the Wigner function as
 \begin{eqnarray}W(q,p)=\frac{1}{2\pi}\iiint\limits_{-\infty}^{\infty}\mathcal{W}(X,\mu,\nu)e^{i(X-\mu q-\nu p)}dX d\mu d\nu.\label{21_5}\end{eqnarray}
\begin{rem} According to the Fubini's theorem the two repeated integrals of a function $f(x,y)$ of two variables are equal if this function is integrable on $X\times Y$, i.e.,
   \begin{eqnarray*}\int_{X}\left(\int_{Y}f(x,y)dx\right)dy=\int_{Y}\left(\int_{X}f(x,y)dy\right)dx.\end{eqnarray*}
Hence, one  may change the order of integration in \eqref{21_4} if $e^{ik(X-\mu q-\nu p)}W(q,p)$ are integrable by $k$, $q$ and $p$
and we may rewrite \eqref{21_4} in the well known form (see \cite{ManciniMankoTombesi})
   \begin{eqnarray*}\mathcal{W}(X,\mu,\nu)=\iiint\limits_{-\infty}^{\infty} e^{ik(X-\mu q-\nu p)}W(q,p)\frac{dkdqdp}{(2\pi)^2}.\end{eqnarray*}
One  may select any integration order. \end{rem}
Obviously the tomogram can be also described in terms of the wave function $\Psi(y)$ by the following relation
   \begin{eqnarray}\mathcal{W}(X,\mu,\nu)=\frac{1}{2\pi|\nu|}\Bigg|\int \Psi(y)e^{\frac{i\mu}{2\nu}y^2-\frac{iX}{\nu}y}dy\Bigg|^2.\label{21_42}\end{eqnarray}
The density matrix of the pure state $\rho_{\Psi}(x,x')$ in the position representation is expressed in terms of the marginal distribution as (see \cite{Mendes})
   \begin{eqnarray}\rho(x,x')=\frac{1}{2\pi}\iint\limits \mathcal{W}(X,\mu,x-x')e^{i(X-\mu(x+x')/2)}dXd\mu.\label{21_10}\end{eqnarray}
 Thus, using the wave function one can reconstruct the marginal distribution in view \eqref{21_42}. And vice versa, if one knows the marginal distribution, the  wave function can also be reconstructed in view of the relationship \eqref{21_10}. In the next sections we consider three special quantum states. The distinctive feature is that their Wigner functions and the tomograms do not satisfy the normalization condition \eqref{21_19} and \eqref{21_20}, respectively.  Therefore, we investigate how the latter fact reflects on the properties of the Wigner function and the tomogram.
 \section{Examples of unnormalized tomograms and  quasi-distributions}\label{sec:21_2}
 \subsection{Plane wave}
\par Let us start from the de Broglie plane wave. The corresponding  wave function in the momentum representation is
 \begin{eqnarray}\Psi_{p}(x)=e^{ip_0x}/\sqrt{2\pi},\label{21_9}\end{eqnarray}
 where $p_0$ is the expectation value of the wave packet's momentum with the center $x_0$.
According to the definition of the density matrix we can write
  \begin{eqnarray}\rho_{p}(x,x')=\Psi_{p}(x)\Psi_{p}^{*}(x')=e^{ip_0(x-x')}/2\pi.\label{21_133}\end{eqnarray}
 Using \eqref{21_2} and the definition of the delta function \eqref{21_28}
 the Wigner function for the latter state can be written as
   \begin{eqnarray}W(q,p)&=&\int\Psi_{p}\left(q+\frac{u}{2}\right)\Psi_{p}^{*}\left(q-\frac{u}{2}\right)e^{-ipu}du\\\nonumber
   &=&\frac{1}{2\pi}\int e^{ip_0\left(q+\frac{u}{2}\right)} e^{-ip_0\left(q-\frac{u}{2}\right)}e^{-ipu}du=\delta(p-p_0).\label{211_155}\end{eqnarray}
It is routine to verify that using \eqref{211_155} the density \eqref{21_133} can be immediately obtained.
Note that the Wigner function \eqref{211_155} does not satisfy the normalization condition \eqref{21_19}, i.e. the integral
   \begin{eqnarray*}\frac{1}{2\pi}\iint\limits_{-\infty}^{\infty}W(q,p)dp dq&=&\frac{1}{2\pi}\iint\limits_{-\infty}^{\infty}\delta(p-p_0)dp dq\end{eqnarray*}
   is not converged. Since such Wigner function is the delta function one can not use the Radon transform  \eqref{21_17} to find the tomogram.
\par Since delta functions are distributions, one have to be careful with verifying whether the usual manipulations are valid. The space of test functions must be strictly specified and the convergence must be checked. Despite it is impossible to define the product at the whole topological vector space of distributions, some distributions may nevertheless be multiplied. A deeper explanation of this phenomenon needs the concept of wavefront sets. Also J.F. Colombeau has developed a theory where the multiplication of distributions is possible \cite{Colombeau,Colombeau2}.
We avoid theoretical details, since it requires substantial knowledge of functional analysis and goes beyond the scope of the article. The basic concepts is that  the Dirac delta function  can be visualized as the limit of a sequence of smooth functions, normalized each to integral equal to one.
\par Let us use the following limit
\begin{eqnarray}\delta(x)=\lim\limits_{\varepsilon\rightarrow 0}\left(\frac{1}{\sqrt{2\pi}\varepsilon}\exp\left(\frac{-x^2}{2\varepsilon^2}\right)\right).\label{21_33}\end{eqnarray}
We substitute \eqref{21_33} instead of the Wigner function in the Radon transform  \eqref{21_4} to find the tomogram of the quantum state \eqref{21_9}.
 After the simple integration one can obtain the tomogram in the explicit form
     \begin{eqnarray}\mathcal{W}_P(X,\mu,\nu)=(2\pi|\mu|)^{-1}.\label{21_122}\end{eqnarray}
Evidently that it  does not also fulfill the normalization condition and it  does not depend on $X$ and $\nu$.
Let us use  $\mathcal{W}_P(X,\mu,\nu)$ in the reversed transformation \eqref{21_10}, namely,
    \begin{eqnarray*}\rho_{p}(x,x')=\!\int \frac{1}{2\pi|\mu|}e^{i(X-\mu(x+x')/2)}dXd\mu
    \neq\frac{1}{2\pi}e^{ip_0(x-x')}.\end{eqnarray*}
    Thus, the tomogram in the form \eqref{21_122} that is independent of the parameters $X$ and $\nu$ does not reconstruct the initial state.
One can assume that in order  to return the initial state the tomogram must depend on all three parameters $(X,\mu,nu)$. Therefore, let us use the Radon transformation in the form
\eqref{21_4} and integrate it by $q$ and $p$, namely
\begin{eqnarray}\mathcal{W}_{P_{int}}(X,\mu,\nu)=\!\int\delta(p-p_0)e^{ik(X-\mu q-\nu p)}\frac{dkdqdp}{(2\pi)^2}=\frac{1}{2\pi}\int\delta(k\mu)e^{ik(X-\nu p_0)}dk.\label{21_11}\end{eqnarray}
One can see that in contrast to \eqref{21_122} the tomogram in the form \eqref{21_11}  contains $X$, $\mu$ and $\nu$. Then using it one can successfully  reconstruct the state, i.e.
   \begin{eqnarray*}\rho_{p}(x,x')&=&\int\left(\delta(k\mu)e^{ik(X-(x-x') p_0}dk\right)e^{i(X-\mu(x+x')/2)}dXd\mu\\
   &=&\frac{1}{2\pi}\int\left(\delta(k\mu)\delta(k+1)e^{-ik(x-x') p_0}dk\right)e^{-i\mu(x+x')/2}d\mu\\
   &=&\frac{1}{2\pi}\int\frac{1}{|k|}\delta(k+1)e^{-ik(x-x') p_0}dk=\frac{1}{2\pi}e^{ip_0(x-x')}.\end{eqnarray*}
However, the delta function in \eqref{21_11} can be viewed as the limit of the sequence of functions \eqref{21_33}.
Substituting the latter limit in \eqref{21_11}, one can write the explicit form of the tomogram \eqref{21_11} in view of a limit
   \begin{eqnarray}\mathcal{W}_{P_{int}}(X,\mu,\nu)=\lim\limits_{\varepsilon\rightarrow 0}\frac{1}{2\pi|\mu|}\exp\left(\frac{-\varepsilon^2(X-\nu p_0)^2}{2\mu^2}\right).\label{21_188}\end{eqnarray}
   Obviously the latter limit coincides with \eqref{21_122}. However, it  contains explicitly all the variables $X$, $\mu$ and $\nu$ on which the tomogram depends.
   Let us use \eqref{21_188} in the reversed transformation \eqref{21_10}, i.e.
    \begin{eqnarray*}\rho_{p}(x,x')&=&\lim\limits_{\varepsilon\rightarrow 0}\frac{1}{2\pi}\int\left(\left(\frac{1}{\sqrt{2\pi}\varepsilon}\exp\left(\frac{-x^2}{2\varepsilon^2}\right)\right)\delta(k+1)e^{-ik(x-x') p_0}dk\right)e^{-i\mu(x+x')/2}d\mu\\
   &=&\frac{1}{(2\pi)^{3/2}}\lim\limits_{\varepsilon\rightarrow 0}\int\frac{1}{\varepsilon}\exp\left(\frac{-\mu^2}{2\varepsilon^2+ip_0(x-x')}\right)e^{-i\mu(x+x')/2}d\mu\\
   &=&\frac{1}{2\pi}\lim\limits_{\varepsilon\rightarrow 0}\exp\left(\frac{-\varepsilon^2(x+x')^2}{2}\right)=\frac{1}{2\pi}e^{ip_0(x-x')}.\end{eqnarray*}
\begin{rem}Another approach can be done using the definition of the tomogram in terms of the wave function \eqref{21_42}. Although we have assumed that $p_0$ in \eqref{21_9} is real, all the above expressions remain valid for complex value $p_0=p_1+ip_2$ until $\mathcal{I}mp_0 < 0$. In this case we have the alternative
representation of the tomogram as
   \begin{eqnarray*}\mathcal{W}_{p_{com}}(X,\mu,\nu)=\frac{1}{2\pi|\nu|}\Bigg|\int \frac{1}{\sqrt{2\pi}}e^{i(p_1+ip_2)y}e^{\frac{i\mu}{2\nu}y^2-\frac{iX}{\nu}y}dy\Bigg|^2
   =\frac{\exp\left(\frac{2p_2}{\mu}(p_1\nu-X)\right)}{2\pi|\mu|}.\end{eqnarray*}
The latter tomogram contains explicitly  $X$, $\mu$ and $\nu$. Then the reversed transformation
 \begin{eqnarray*}\rho_{p_{com}}(x,x')&=&\frac{1}{2\pi}e^{ip_1(x-x')}e^{-ip_2(x-x')}\end{eqnarray*}
 can be done using \eqref{21_10}.
Since $p_2=0$ and $p_1\equiv p_0$ hold, the latter result coincides with \eqref{21_133}.
\end{rem}
\par The $Q$-function for the latter state can be easily found using \eqref{21_8}, e.g.,
    \begin{eqnarray*}Q(q,p,t)=\frac{1}{\pi}\int \delta(p'-p_0)e^{-(q'-q)^2}e^{-(p'-p)^2}dq'dp'=\frac{1}{\sqrt{\pi}}e^{-(p-p_0)^2}.\\\end{eqnarray*}
    Obviously it does not fulfil the normalization condition too.
\par On the other hand, we can take a state with a wave function
  \begin{eqnarray}\Psi_{x}(x)=\delta(x-x_0).\label{21_31}\end{eqnarray}
  The density matrix is in this case the following
  \begin{eqnarray}\rho_{x}(x,x')=\Psi_{x}(x)\Psi_{x}^{*}(x')=\delta(x-x_0)\delta(x'-x_0).\label{20_13}\end{eqnarray}
It is easy to verify that its Wigner function and the $Q$-function are
   \begin{eqnarray}&&W(q,p)=\delta(x-x_0),\quad Q(q,p)=\frac{1}{\sqrt{\pi}}e^{-(x-x_0)^2}.\label{20_32}\end{eqnarray}
Similarly to the previous example the tomogram can be written as
   \begin{eqnarray}&&\mathcal{W}_1(X,\mu,\nu)=\frac{1}{2\pi}\int\delta(k\nu)e^{ik(X-\mu x_0)}dk.\label{20_33}\end{eqnarray}
   Using the technics described above we can replace the delta function in \eqref{20_33} by its limit representation \eqref{21_33} and reconstruct the density matrix.
Thus, it was shown by the examples that the tomogram  must explicitly contain all variables on which the tomogram depends, i.e., $X$, $\mu$ and $\nu$, to reconstruct the state.
 For the case of a plane wave this can be achieved by using the limiting expression for the delta function. Note that the plane wave state  is special since the Wigner function is the Dirac delta function, so one has to be careful with the verifying that the usual manipulations are valid.
 The space of test functions must be correctly indicated and the convergence must be checked.
 \subsection{Moshinsky shutter}
 The diffraction in time phenomena was introduced in \cite{Moshinsky}. A stream of particles of $m=1$, $\hbar=1$ of energy $k^2/2$ moves parallel to the $x$-axes. It is interrupted by a completely absorbing shutter situated at $x=0$ which is opened at time $t=0$. The problem is to find $\psi(x,t)$ that satisfies the free one dimensional time dependent Schrodinger equation, i.e.,
\begin{eqnarray*}i\frac{\partial\psi(x,t)}{\partial t}=-\frac{1}{2}\frac{\partial^2\psi(x,t)}{\partial x^2}.\end{eqnarray*}
The initial condition is
\begin{eqnarray*}\psi(x,0)=e^{ikx}\theta(-x),\end{eqnarray*}
where $\theta(x)$ is a step function defined as
\begin{equation*}
    \begin{matrix}
    \theta(x) & =
    & \left\{
    \begin{matrix}
    1 &  \mbox{if}\quad x>0\\
   0 & \mbox{if}\quad x<0
    \end{matrix} \right.
    \end{matrix}.
\end{equation*}
Here $k$  is assumed to be real.
The solution of this problem reads as (see \cite{MankoMoshinskySharma,Nussenzveig,Moshinsky})
\begin{eqnarray*}M(x,k,t)=\frac{1}{2}\exp\left(i\left(kx-\frac{1}{2}k^2t\right)\right)\operatorname{erfc}\left(e^{-i\pi/4}\omega\right),\end{eqnarray*}
where $\omega=\frac{x-kt}{\sqrt{2t}}$  and the error integral is
\begin{eqnarray*}\operatorname{erfc}\left(z\right)=\frac{2}{\sqrt{\pi}}\int_{z}^{\infty}e^{-y^2}dy.\end{eqnarray*}
In \cite{Moshinsky} the latter solution was rewritten in the following form
\begin{eqnarray*}&M(x,k,t)=e^{-i\pi/4}e^{i\left(kx-\frac{1}{2}k^2t\right)}\frac{1}{\sqrt{2}}\left(\Big[\frac{1}{2}-C(\omega)\Big]+i\Big[\frac{1}{2}-S(\omega)\Big]\right),\end{eqnarray*}
where the Fresnel integrals were defined as
\begin{eqnarray}C(\omega)=\sqrt{\frac{2}{\pi}}\int_{0}^{\omega}\cos y^2dy,\quad S(\omega)=\sqrt{\frac{2}{\pi}}\int_{0}^{\omega}\sin y^2dy.\label{20_17}\end{eqnarray}
It is also straightforward to verify that the density matrix is
\begin{eqnarray}\rho(x,x',k,t)=\frac{1}{4}\exp\left(ik\left(x-x'\right)\right)\operatorname{erfc}\left(e^{-i\pi/4}\omega\right)\operatorname{erfc}\left(e^{i\pi/4}\omega'\right).\label{21_21}\end{eqnarray}
The Wigner function for the diffraction in time problem was obtained in \cite{MankoMoshinskySharma} and has the following representation
\begin{eqnarray*}W(x,p,k,t)=-\frac{1}{2\pi^2}e^{2ip(x-pt)}\int\limits_{-\infty}^{+\infty}\frac{e^{-2i\kappa(x-pt)}}{(\kappa-k)(\kappa+k-2p)}d\kappa\end{eqnarray*}
or in the explicit form as
\begin{eqnarray}W(x,p,k,t)=\frac{\theta(pt-x)}{4\pi i(k-p)}\left(e^{2i(k-p)(x-pt)}-e^{-2i(k-p)(x-pt)}\right).\label{21_34}\end{eqnarray}
Substituting this in the normalization condition  \eqref{21_19} one can obtain
\begin{eqnarray*}&-&\frac{1}{4\pi^3}\iiint\limits_{-\infty}^{\infty}\frac{e^{2ix(p-\kappa)-2ip^2t+2i\kappa pt}}{(\kappa-k)(\kappa+k-2p)}d\kappa dpdx=
-\frac{1}{4\pi^2}\iint\limits_{-\infty}^{\infty}\frac{\delta(p-\kappa)e^{-2ip^2t+2i\kappa pt}}{(\kappa-k)(\kappa+k-2p)}d\kappa dp\\
&=&\frac{1}{4\pi^2}\int\limits_{-\infty}^{\infty}\frac{1}{(\kappa-k)^2}d\kappa =0.\end{eqnarray*}
Thus, the normalization condition is violated. The tomogram was found in \cite{MankoMoshinskySharma} by the wave function and has the form
\begin{eqnarray}\mathcal{W}(X,\mu,\nu,k,t)&=&\frac{1}{2|\mu|} \left(\Bigg(\frac{1}{2}+C(\rho)\Bigg)^2+\Bigg(\frac{1}{2}+S(\rho)\Bigg)^2\right),\\
\rho&=&\frac{k(\mu t+\nu)-X)}{\sqrt{2\mu(\mu t+\nu)}},\label{21_22}\end{eqnarray}
where $C(\cdot)$ and $S(\cdot)$ are the Fresnel integrals \eqref{20_17}.
As in the previous example, one can see that for the diffraction in time problem neither the Wigner function nor the tomogram satisfy the normalization conditions.
However, unlike a plane wave, the tomogram  depends directly on $X$, $\mu$ and $\nu$. Thus, it is interesting to do the transition \eqref{21_4} explicitly  from the Wigner function to the tomogram that was not directly done before, as well as the reconstruction of the state by the Wigner function and tomogram. All the transformations are done in Appendix \ref{ap_2}.

  \subsection{The charge moving in homogeneous electric field}
  In this section, we turn our attention to the problem of the charged particle moving in the uniform and constant electric field. The Wigner and the tomographic descriptions of the state are done in
  \cite{shchukin}. The Schr\"{o}dinger equation for the stationary state of the particle in the electric field
  reads as
  \begin{eqnarray*}\frac{d^2\psi}{dx^2}+\frac{2m}{\hbar^2}(\mathcal{E}+\mathcal{F}x)\psi=0, \end{eqnarray*}
  where for all the energy values $\mathcal{E}$ there exists the following solution
   \begin{eqnarray*} \psi_{\mathcal{E}}(x)
   &=&A\Phi(\alpha x+\varepsilon).\end{eqnarray*}
 Here, the following notations were used
     \begin{eqnarray*}A=\frac{(2m)^{1/3}}{\pi^{1/2}\mathcal{F}^{1/6}\hbar^{2/3}},\quad \alpha=-\left(\frac{2m\mathcal{F}}{\hbar^2}\right)^{1/3},\quad \varepsilon=\alpha\frac{\mathcal{E}}{\mathcal{F}}. \end{eqnarray*}
   Here $\mathcal{F}$ is a uniform electric field.
  The Airy function of the first kind $\operatorname{Ai}(x)=\Phi(x)/\sqrt{\pi}$ is bounded on $\mathbb{R}$ and has the integral representation for $x\in\mathbb{R}$
     \begin{eqnarray*}\operatorname{Ai}(x)&=&\frac{1}{\pi}\int\limits_{0}^{\infty}\cos\left(x\xi+\frac{\xi^3}{3}\right)d\xi. 
     \end{eqnarray*}
     According to \cite{Soares} the product of two Airy functions reads as
           \begin{eqnarray} \operatorname{Ai}(x)\operatorname{Ai}(y)&=&\frac{1}{2\pi^{3/2}}\int\limits_{0}^{\infty}
           \frac{\cos\left(\frac{x+y}{2}\xi+\frac{\xi^3}{12}-\frac{(x-y)^2}{4\xi}+\frac{\pi}{4}\right)}{\sqrt{\xi}}d\xi.\label{21_37}\end{eqnarray}
Hence, the density matrix can be written as
 \begin{eqnarray}\label{21_36}\rho(x,x')&=& A^2\Phi(\alpha x+\varepsilon)\Phi^{\dagger}(\alpha x'+\varepsilon)\\\nonumber
 &=&\frac{A^2}{2\pi^{3/2}}\int\limits_{0}^{\infty}\frac{\cos\left(\frac{\alpha(x+x')+2\varepsilon}{2}\xi+\frac{\xi^3}{12}-\frac{\alpha^2(x-x')^2}{4\xi}+\frac{\pi}{4}\right)}{\sqrt{\xi}}d\xi.
 \end{eqnarray}
     The Wigner function of the stationary state of a charged particle with a given energy is
               \begin{eqnarray*} W_{\mathcal{E}}(q,p)=\frac{1}{\pi}\sqrt[3]{\frac{m}{\mathcal{F}^2\hbar^2}}\Phi\left(\sqrt[3]{4}\left(\frac{p^2}{\alpha^2\hbar^2}+\alpha q+\varepsilon\right)\right)
               \end{eqnarray*}
               and the tomogram is
 \begin{eqnarray} W_{\mathcal{E}}(X,\mu,\nu)=\frac{A^2\hbar}{4\pi|\mu|}\Bigg|\Phi\left(\varepsilon-\frac{\alpha X}{\mu}-\frac{\hbar^2\alpha^2\nu^2}{4\mu^2}\right)\Bigg|^2.\label{21_35}\end{eqnarray}
 The following representation holds for $x,a,b\in \mathbb{R}$ (see \cite{Dominici,Soares2})
 \begin{eqnarray*}\int\limits_{-\infty}^{\infty}\operatorname{Ai}(x-a)\operatorname{Ai}(x-b)dx=\delta(a-b). \end{eqnarray*}
 Hence, the normalization condition for the tomogram $ W_{\mathcal{E}}(X,\mu,\nu)$ is
\begin{eqnarray*}\frac{A^2\hbar}{4|\mu|} \int\limits_{-\infty}^{\infty}\Bigg|\operatorname{Ai}\left(\varepsilon-\frac{\alpha X}{\mu}-\frac{\hbar^2\alpha^2\nu^2}{4\mu^2}\right)\Bigg|^2dX
\end{eqnarray*}
that does not converge.
The tomogram \eqref{21_35} depends on all the variables, $X$, $\mu$ and $\nu$, and reconstructs the initial state. The details are given in Appendix \ref{ap_1}.
\section{Conclusion}
We have studied a special class of quantum states whose  tomogram and Wigner, $P$- and $Q$- functions  do not satisfy the standard normalization condition \eqref{21_01}. The special conditions under which the latter functions do not reconstruct the initial state are studied in details. In particular, it is shown that if the Wigner function contains the  Dirac delta function, then in general the classical Radon transform \eqref{21_16} can not be applied. This means that for such class of the Wigner functions it is impossible to determine the tomogram. Authors have shown that one can overcame this problem by using the limiting relation of the delta function. The effectiveness of this approach is illustrated by the example of the plane wave. The corresponding Wigner function is the delta function and obviously it does not satisfy the normalization condition. Next, it is shown that the Wigner function or the tomogram can violate the normalization condition, but if they are continuous functions of all their variables, they reconstruct still  the density matrix. This fact is illustrated by two examples, namely, by the Moshinsky shutter problem and by the motion of a charge in homogeneous electric field. Their quasi-distributions and tomograms have complex structures and depend on the Airy functions and the Fresnel integrals, but they contain all the necessary variables in the explicit form and are continuous functions. In this case, all the inverse transforms that reconstruct the density matrix or the wave function are applicable without limitations.
Thus, the normalization condition is not a determining factor. The continuity and the dependence on all of necessary variables in an explicit form of the quasi-distributions and the tomogram are the most important.
\appendix
\section{Appendix}
 \subsection{Transition from Wigner function for Moshinsky shutter to tomogram}\label{ap_2}
 Using the Radon transform of the Wigner function \eqref{21_4} one can rewrite the tomogram for the diffraction in time case as
   \begin{eqnarray*}\mathcal{W}(X,\mu,\nu,k,t)=\iiint\limits_{-\infty}^{\infty} e^{iz(X-\mu x-\nu p)}\frac{\theta(pt-x)}{16\pi^3 i(k-p)}\left(e^{2i(k-p)(x-pt)}-e^{-2i(k-p)(x-pt)}\right)dxdpdz.\end{eqnarray*}
Replacing variables by
 \begin{eqnarray}u\equiv pt-x,\quad v\equiv pt+x\label{21_29}\end{eqnarray}
and using the relation
\begin{eqnarray}\int_{0}^{\infty}e^{-iyz}dy=-\frac{i}{z}, \label{21_23}\end{eqnarray}
one can rewrite the latter tomogram as
 \begin{eqnarray*}\mathcal{W}(X,\mu,\nu,k,t)&=&\frac{1}{32\pi^3t}\iint\limits_{-\infty}^{\infty} \iint_{0}^{\infty}\Bigg(e^{ui\frac{2v+y-(\nu+\mu t)z}{2t}
 -vi\left(2k-\frac{v}{t}\right)-yi\left(k-\frac{v}{2t}\right)
 +zi\left(X+v\left(\frac{\mu}{2}-\frac{\nu}{t}\right)\right)}\nonumber\\
 &-&e^{-ui\frac{2v-y+(\nu+\mu t)z}{2t}
 +vi\left(2k-\frac{v}{t}\right)-yi\left(k-\frac{v}{2t}\right)
 +zi\left(X+v\left(\frac{\mu}{2}-\frac{\nu}{t}\right)\right)}\Bigg)dudzdydv.\end{eqnarray*}
Using the definition of the delta function \eqref{21_28} one can rewrite both integrals by $u$ in the last formula can be written as
 \begin{eqnarray*}&&\int\limits_{-\infty}^{\infty}e^{\pm ui\frac{2v\pm y\mp(\nu+\mu t)z}{1}}du=\frac{2t}{|\nu+\mu t|}\delta\left(z\mp\frac{2v\pm y}{\nu+\mu t}\right).
 \end{eqnarray*}
Next, integrating by $z$ one can write
 \begin{eqnarray}\mathcal{W}(X,\mu,\nu,k,t)&=&\frac{t}{4\pi^2|\nu+\mu t|} \iint_{0}^{\infty}
 \Bigg(e^{\frac{2i\mu }{\nu+\mu t}v^2+\frac{i(2X+\mu y-2k(\nu+\mu t))}{\nu+\mu t}v-\frac{iy(k(\nu+\mu t)-X)}{\nu+\mu t}}\\\nonumber
 &-&e^{-\frac{2i\mu }{\nu+\mu t}v^2+\frac{i(-2X+\mu y+2k(\nu+\mu t))}{\nu+\mu t}v-\frac{iy(k(\nu+\mu t)-X)}{\nu+\mu t}}\Bigg)dydv.\label{21_26}\end{eqnarray}
 According to the known formula (see \cite{Edward})
 \begin{eqnarray}\int_{\omega}^{\infty}e^{-(ax^2+bx+c)}dx&=&\frac{1}{2}\sqrt{\frac{\pi}{a}}e^{\frac{b^2-4ac}{4a}}
 \left(1-\operatorname{erf}\left(\omega\sqrt{a}+\frac{b}{2\sqrt{a}}\right)\right)\label{21_24}
 \end{eqnarray}
  the integration by $v$ gives the following result
  \begin{eqnarray*}\mathcal{W}(X,\mu,\nu,k,t)&=&\frac{1}{8\pi\sqrt{\pi}\sqrt{2\mu(\nu+\mu t)}} \\
 &\cdot&\int_{0}^{\infty}\Bigg(\sqrt{i}e^{-i\left(\rho+\frac{\mu y}{2\sqrt{2\mu(\nu+\mu t)}}\right)^2}\operatorname{erfc}\left(\frac{i}{\sqrt{-i}}\left(\frac{\mu y}{2\sqrt{2\mu(\nu+\mu t)}}-\rho\right) \right)\\
 &-&\sqrt{-i}e^{i\left(\frac{\mu y}{2\sqrt{2\mu(\nu+\mu t)}}-\rho\right)^2}\operatorname{erfc}\left(\frac{-i}{\sqrt{i}}\left(\rho+\frac{\mu y}{2\sqrt{2\mu(\nu+\mu t)}}\right) \right)\Bigg)dy.\end{eqnarray*}
Replacing the variables, one can write
   \begin{eqnarray*}\mathcal{W}(X,\mu,\nu,k,t)&=&\frac{1}{4\mu\pi\sqrt{\pi}}
 \Bigg(\int_{-\rho}^{\infty}\sqrt{i}e^{-i\left(x_1+2\rho\right)^2}\operatorname{erfc}\left(\frac{i}{\sqrt{-i}}x_1 \right)dx_1\\
 &-&\int_{\rho}^{\infty}\sqrt{-i}e^{i\left(x_2-2\rho\right)^2}\operatorname{erfc}\left(\frac{-i}{\sqrt{i}}x_2 \right)\Bigg)dx_2.\end{eqnarray*}
 Using the integration by parts, where
   \begin{eqnarray*}&&u_1= \operatorname{erfc}\left(\frac{i}{\sqrt{-i}}x_1\right),\quad v_1=-\frac{\sqrt{\pi}}{2}(-1)^{3/4}\operatorname{erf}\left((x_1+2\rho)(-1)^{1/4}\right),\\
   \end{eqnarray*}
      \begin{eqnarray*}&&u_2= \operatorname{erfc}\left(\frac{-i}{\sqrt{i}}x_2\right),\quad v_2=-\frac{\sqrt{\pi}}{2}(-1)^{1/4}\operatorname{erf}\left((x_2-2\rho)(-1)^{3/4}\right),\\
  \end{eqnarray*}
 the latter integral follows
   \begin{eqnarray*}\mathcal{W}(X,\mu,\nu,k,t)&=&\frac{1}{8|\mu|\pi} \Bigg((1 - \operatorname{erf}((-1)^{3/4} x_1)) \operatorname{erf}((-1)^{1/4} (2 \rho + x_1))\Bigg|_{-\rho}^{\infty}\\
   &+&  (i-1) \sqrt{\frac{2}{\pi}}\int_{-\rho}^{\infty}\operatorname{erf}((x_1+2\rho)(-1)^{1/4})e^{ix_1^2}dx_1\\
 &+&(\operatorname{erf}((-1)^{1/4} x_2) + 1) \operatorname{erf}((-1)^{3/4} (x_2 - 2 \rho))\Bigg|_{\rho}^{\infty}\\
 &-&(1+i) \sqrt{\frac{2}{\pi}}\int_{\rho}^{\infty}\operatorname{erf}((x_2-2\rho)(-1)^{3/4})e^{-ix_2^2}dx_2\Bigg).
\end{eqnarray*}
  Hence, after the transformation, we cam obtain
 \begin{eqnarray}\mathcal{W}(X,\mu,\nu,k,t)=\frac{1}{4|\mu|} \Bigg(1+
 \frac{1}{\sqrt{2}}\left( \operatorname{erf}(\sqrt{i}\,\rho) + \operatorname{erf}(\sqrt{-i}\,\rho) \right)
 +\operatorname{erf}(\sqrt{i}\,\rho)\operatorname{erf}(\sqrt{-i}\,\rho)\Bigg).\label{21_25}\end{eqnarray}
 Using the known expressions
 \begin{eqnarray*}S(x)&=&\frac{1}{2\sqrt{2}} \left( \sqrt{i}\,\operatorname{erf}(\sqrt{i}\,x) + \sqrt{-i}\,\operatorname{erf}(\sqrt{-i}\,x) \right),\\
 C(x)&=&\frac{1}{2\sqrt{2}} \left( \sqrt{-i}\,\operatorname{erf}(\sqrt{i}\,x) + \sqrt{i}\,\operatorname{erf}(\sqrt{-i}\,x) \right)\end{eqnarray*}
of the Fresnel integrals \eqref{20_17} in terms of the error function, the latter tomogram can be rewritten as \eqref{21_22}.
\subsection{Reconstruction of Moshinsky shutter state from Wigner function}
    Let us use the Wigner function \eqref{21_34} in the reversed transformation \eqref{21_3}, i.e.
\begin{eqnarray*}\rho(x,x',k,t)=\frac{1}{2\pi}\int\limits_{-\infty}^{\infty} \frac{\theta\left(pt-\frac{x+x'}{2}\right)}{4\pi i(k-p)}\left(e^{2i(k-p)(\frac{x+x'}{2}-pt)}-e^{-2i(k-p)(\frac{x+x'}{2}-pt)}\right) e^{ip(x-x')}dp.\end{eqnarray*}
Changing the variables as $P=pt-\frac{x+x'}{2}$, one can write
\begin{eqnarray*}\rho(x,x',k,t)=\frac{1}{8\pi^2i}\int\limits_{-\infty}^{\infty} \frac{\theta\left(P\right)e^{i(x-x')\left(\frac{p}{t}+\frac{x+x'}{2t}\right)}}{k-\left(\frac{p}{t}+\frac{x+x'}{2t}\right)}\left(e^{-2iP\left(k-\left(\frac{p}{t}+\frac{x+x'}{2t}\right)\right)}-e^{2iP(\left(k-\left(\frac{p}{t}+\frac{x+x'}{2t}\right)\right)}\right) dP.\end{eqnarray*}
Using notations $\omega\equiv\frac{x-kt}{\sqrt{2t}}$, $\omega'\equiv\frac{x'-kt}{\sqrt{2t}}$ one can rewrite the latter integral in the following form
\begin{eqnarray*}\rho(x,x',k,t)=\frac{i\sqrt{2}e^{ik(x-x')}}{8\sqrt{t}\pi^2}\int\limits_{0}^{\infty}\frac{e^{i(\sqrt{\frac{2}{t}}P+\omega+\omega')(\sqrt{\frac{2}{t}}P+\omega-\omega')}
-e^{-i(\sqrt{\frac{2}{t}}P+\omega+\omega')(\sqrt{\frac{2}{t}}P-\omega+\omega')}}{\sqrt{\frac{2}{t}}P+\omega+\omega'}
 dP.\end{eqnarray*}
Let us denote $x_1\equiv\sqrt{\frac{2}{t}}P+\omega$, $x_2\equiv\sqrt{\frac{2}{t}}P+\omega'$. Then, we can rewrite the density matrix as
\begin{eqnarray*}\rho(x,x',k,t)=\frac{ie^{ik(x-x')}}{8\pi^2}\Bigg[\int\limits_{\omega}^{\infty}\frac{e^{i(x_1^2-\omega'^2)}}{x_1+\omega'}dx_1-
\int\limits_{\omega'}^{\infty}\frac{e^{-i(x_2^2-\omega^2)}}{x_2+\omega}dx_2\Bigg].
 \end{eqnarray*}
Taking into account the formula \eqref{21_23} 
and using again the relation \eqref{21_24} we write
 \begin{eqnarray}\rho(x,x',k,t)&=&\frac{-e^{ik(x-x')}}{8\pi^2}\Bigg[\int_{0}^{\infty}\frac{1}{2}\sqrt{\frac{\pi}{-i}}
 e^{\frac{-i(y+2\omega')^2}{4}}\operatorname{erfc}\left(\frac{(y-2\omega)i}{2\sqrt{-i}}\right)dy\nonumber\\
 &-&\int_{0}^{\infty}\frac{1}{2}\sqrt{\frac{\pi}{i}}
 e^{\frac{i(y-2\omega)^2}{4}}\operatorname{erfc}\left(\frac{(y+2\omega')i}{2\sqrt{i}}\right)dy\Bigg].\label{21_27}
 \end{eqnarray}
 Finally, using integration by parts as it was done in the previous example the latter integral leads to \eqref{21_21}.
 \subsection{Reconstruction of Moshinsky shutter state from tomogram}
  We begin from the tomogram \eqref{21_22} in the form \eqref{21_26}.
  Using  \eqref{21_10} one can write
  \begin{eqnarray*}\rho(x,x',k,t)&=&\frac{t}{8\pi^3}\iint_{-\infty}^{\infty} \frac{1}{|x-x'+\mu t|}\\
  &\cdot& \iint_{0}^{\infty}
 \Bigg(e^{\frac{2i\mu }{x-x'+\mu t}v^2+\frac{i(\mu y-2k(x-x'+\mu t))}{x-x'+\mu t}v-iyk}e^{iX\left(\frac{2v+y}{x-x'+\mu t}+1\right)}\\
 &-&e^{-\frac{2i\mu }{x-x'+\mu t}v^2+\frac{i(\mu y+2k(x-x'+\mu t))}{x-x'+\mu t}v-iyk}e^{iX\left(\frac{-2v+y}{x-x'+\mu t}+1\right)}\Bigg)dydve^{-i\mu(x+x')/2)}dXd\mu.\end{eqnarray*}
The integrals depending on $X$
     \begin{eqnarray*}\frac{1}{2\pi}\int_{-\infty}^{\infty}e^{\pm iX\left(\frac{\pm2v+y}{x-x'+\mu t}+1\right)}dX=\frac{|x-x'+\mu t|}{|t|}\delta\left(\mu-\frac{x'-x-y\mp2v}{t} \right)\end{eqnarray*}
hold. Moreover, integrating the density by $\mu$, one can write
  \begin{eqnarray*}\rho(x,x',k,t)&=&\frac{1}{4\pi^2}  \iint_{0}^{\infty}
 \Bigg(e^{i\left(\frac{2v^2}{t}+\frac{2v(x-kt)+y}{t}+\frac{(x+x'-2kt)y+x^2-x'^2}{2t}\right)}\\
 &-&e^{i\left(\frac{-2v^2}{t}+\frac{-2v(x'-kt)+y}{t}+\frac{(x+x'-2kt)y+x^2-x'^2}{2t}\right)}\Bigg)dydv.\end{eqnarray*}
 Using \eqref{21_24} and  integrating by $v$ one can obtain \eqref{21_27}.
\subsection{Reconstruction of stationary state of a charged particle with a given energy from tomogram}\label{ap_1}
 According to the integral representation of the two Airy functions given by \eqref{21_37}, the tomogram \eqref{21_35}  can be written as follows
   \begin{eqnarray*}  W_{\mathcal{E}}(X,\mu,\nu)=\frac{A^2\hbar}{8\pi^{3/2}|\mu|}\int\limits_{0}^{\infty}\frac{\cos\left(\left(\varepsilon-\frac{\alpha X}{\mu}-\frac{\hbar^2\alpha^2\nu^2}{4\mu^2}\right)\xi+\frac{\xi^3}{12}+\frac{\pi}{4}\right)}{\sqrt{\xi}}d\xi.\end{eqnarray*}
Using the definition \eqref{21_10} and due to the Euiler formula and expression
   \begin{eqnarray*}\int\limits_{-\infty}^{\infty}e^{iX(1\pm\frac{\xi\alpha}{\mu})}dX=\delta\left(1\pm\frac{\xi\alpha}{\mu}\right), \end{eqnarray*}
the density matrix can be reconstructed using  the latter tomogram as
 \begin{eqnarray*}\rho(x,x')&=&\frac{A^2\hbar}{32\pi^{3/2}}\int\limits_{0}^{\infty}\frac{1}{\sqrt{\xi}}d\xi
 \int\limits_{-\infty}^{\infty}\Bigg(\delta(\mu-\xi\alpha)e^{i(\xi(\varepsilon-\frac{\hbar^2\alpha^2(x-x')^2}{4\mu^2}-\frac{\mu(x+x')}{2})+\frac{\xi^3}{12}+\frac{\pi}{4}))}\\
 &+&\delta(\mu+\xi\alpha)e^{-i(\xi(\varepsilon-\frac{\hbar^2\alpha^2(x-x')^2}{4\mu^2}-\frac{\mu(x+x')}{2})+\frac{\xi^3}{12}+\frac{\pi}{4}))}\Bigg)d\mu.
 \end{eqnarray*}
 After the integration by $\mu$ and some simplifications one can obtain the density matrix \eqref{21_36}.

\begin{acknowledgements}The study in section 3 and 4 by Markovich L.A. was supported by the Russian Science Foundation grant (14-50-00150).
\end{acknowledgements}



\end{document}